\documentclass[11pt,a4paper,english,superscriptaddress,prd,aps,preprint]{revtex4}
\usepackage{amssymb}
\usepackage{amsmath} 
\usepackage{slashed}
\usepackage{graphicx}
\usepackage{babel}
\usepackage{hyperref}
\usepackage{color}
\makeatletter
\AtBeginDocument{\let\@elt\relax}\makeatother

\begin{document}

\title{Two-loop corrections to the Carroll-Field-Jackiw term in a CPT-odd Lorentz-violating scalar QED}

\author{A. C. Lehum}
\email{lehum@ufpa.br}
\affiliation{Faculdade de F\'{i}sica, Universidade Federal do Par\'{a}, 66075-110, Bel\'{e}m, Par\'a, Brazil}

\author{J. R. Nascimento}
\email{jroberto@fisica.ufpb.br}
\affiliation{Departamento de F\'{i}sica, Universidade Federal da Para\'{i}ba, Caixa Postal 5008,
	58051-970 Jo\~{a}o Pessoa, Para\'iba, Brazil}
	
\author{A. Yu. Petrov}
\email{petrov@fisica.ufpb.br}
\affiliation{Departamento de F\'{i}sica, Universidade Federal da Para\'{i}ba, Caixa Postal 5008,
	58051-970 Jo\~{a}o Pessoa, Para\'iba, Brazil}

\begin{abstract}
In this study, we systematically calculate one-loop corrections to the Lorentz-violating vertices within the framework of CPT-odd Quantum Electrodynamics, encompassing scalar and photon fields in arbitrary gauge. Additionally, we ascertain the finite two-loop corrections to the Carroll-Field-Jackiw term. Furthermore, we analyze the UV divergent component of the two-loop Lorentz-violating correction in the self-energy of the scalar field.
\end{abstract}

\maketitle

\section{Introduction}

Perturbative calculations in various Lorentz-violating (LV) theories represent an important line of study of these theories. The first examples of such calculations were given already in the seminal papers \cite{ColKost1,ColKost2} where the Lorentz-violating standard model extension (LV SME) has been formulated. The paradigmatic role was played by the paper \cite{JK} where the Carroll-Field-Jackiw (CFJ) term has been shown to arise as a quantum correction in CPT-odd LV spinor QED, giving thus a methodology and a motivation to further studies of perturbative generation of various LV terms depending on scalar and/or gauge fields (see \cite{ourrev} for a review on these studies).  However, all these calculations, including those ones discussed in \cite{ourrev}, were performed only in the one-loop approximation. Therefore, the natural problem consists in studying of possibility of arising new effects in higher loop corrections. Just now, there are only a few examples of higher-loop calculations in LV theories -- for the spinor QED, an interesting example is presented in the paper \cite{noCFJ}, where the absence of higher-loop contributions to the CFJ term has been proved. 

At the same time, it is interesting to study Lorentz symmetry breaking not only in the spinor QED but also in the scalar one, considering the Higgs sector of the LV SME \cite{ColKost2}. Earlier, the CPT-even scalar QED and its non-Abelian generalization have been considered in the one-loop approximation \cite{brito1,brito2,brito3}, and, further, the two-loop contributions in this theory were studied \cite{our2loop}. Also, it is worth to mention a discussion of two-loop corrections in the LV Yukawa theory given in \cite{Yukawa}. Therefore, it is natural to calculate two-loop corrections, for the CPT-odd LV scalar QED which earlier has been studied only in the one-loop approximation within the Higgs mechanism context \cite{Scarp2013}.

Within this paper, we pursue the following aims. For the CPT-odd LV QED defined as a further generalization of that one proposed in \cite{Scarp2013}, we calculate the one-loop contributions to self-energy tensors of scalar and gauge field, and further, the two-loop contribution to the self-energy of the gauge field, in the lower order in the LV parameters $b^{\mu}$ and $u^{\mu}$. 

The structure of the paper looks like follows. 
	In Section 2, we introduce our model, specifically the massless CPT-odd LV scalar QED. Section 3 is dedicated to the one and two-loop calculations. Our summary is provided in Section 4, while the comprehensive details of the calculations can be found in the Supplemental Material.


Throughout this paper, we use natural units $c=\hbar=1$ and $(+---)$ as the spacetime signature. 

\section{The massless CPT-odd Lorentz-violating Scalar electrodynamics}\label{sec01}

Let us consider the model described by the Lagrangian
\begin{eqnarray}\label{eq01}
\mathcal{L}&=&  (1+\delta_2) (D^{\mu}\phi)^\dagger D_{\mu}\phi - \frac{\lambda(1+\delta_\lambda)}{4} (\phi^\dagger\phi)^2+\frac{i}{2}Q_1(1+\delta_{Q_1}) u^{\mu}\left[ \phi^* D_\mu\phi-\phi  (D_{\mu}\phi)^*\right]\nonumber\\
&& -\frac{(1+\delta_3)}{4}F^{\mu\nu}F_{\mu\nu}+\frac{Q_2(1+\delta_{Q_2})}{2}\epsilon^{\mu\nu\alpha\beta}b_{\mu}A_{\nu}F_{\alpha\beta},
\end{eqnarray}
\noindent 
	where $D_{\mu}=\partial_{\mu} - ieA_{\mu}$ represents the covariant derivative, while $u^{\mu}$ is a Lorentz-violating (LV) CPT-odd constant vector field and $b^{\mu}$ is a LV CPT-odd constant pseudovector field. The model involves LV coupling constants, denoted as $Q_1$ and $Q_2$. The coefficients $\delta_2$, $\delta_{3}$, $\delta_\lambda$, $\delta_{Q_1}$, and $\delta_{Q_2}$ correspond to the counterterms for the scalar field wave-function, photon field wave-function, $\lambda$ scalar self-coupling and the LV coupling constants $Q_1$ and $Q_2$, respectively. Actually, this theory represents itself as an extension of the model studied in \cite{Scarp2013}, where an extra term, involving Lorentz symmetry breaking in the scalar sector as well, is added. We note that both these LV terms have been introduced originally within the context of LV SME \cite{ColKost1,ColKost2}.
Here, the $b^{\mu},u^{\mu}$ are assumed to be VEVs of some fields, whose dynamics does not matter within our study. We note that they can be treated as real ones without any difficulties, cf. \cite{ColKost1,ColKost2}.

To consider the theory perturbatively, one must add to its Lagrangian the gauge-fixing term. We choose the usual one, with $\xi$ is the gauge fixing parameter:
\begin{equation}
	{\cal L}_{gf}=-\frac{1}{2\xi}(\partial_{\mu}A^{\mu})^2,
\end{equation}
which yields the standard propagator of the gauge field in an arbitrary gauge. The propagator of the scalar field is also the usual one.

Our primary objective is to explore the ultraviolet (UV) properties of the model. Consequently, we do not address the infrared (IR) divergences that might manifest in the massless model. Typically, in the intermediate phases of the calculation, infrared issues can be circumvented by incorporating a mass regulator in the field propagators.


\section{The radiative corrections for the Lorentz-violating vertices}

In order to perform the calculations for both one and two-loop diagrams, we make use of a customized version of a suite of MATHEMATICA packages~\cite{feyncalc,feynarts,feynrules,Shtabovenko:2016whf}. Specifically, for the two-loop diagrams, we employ the Tarasov algorithm \cite{tarasov} to simplify the two-loop integrals into a set of fundamental ones. The implementation of the Tarasov algorithm is facilitated through the TARCER package \cite{tarcer}, where the basic integrals have been computed as documented in \cite{tsil}. These computational tools greatly enhance the efficiency of calculating and manipulating Feynman diagrams and associated quantities in our analysis. For a comprehensive treatment of the two-loop diagrams, we present a detailed calculation in the Supplemental Material.

\subsection{Calculation of the LV corrections to the one-loop  self-energies} \label{sec2-oneloop}

\begin{figure}[ht]
	\includegraphics[angle=0,width=16cm]{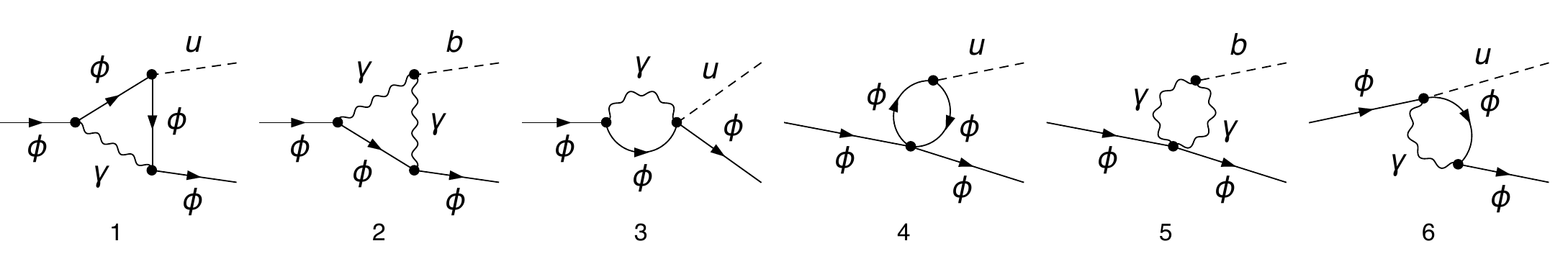}
	\caption{Feynman diagrams for the LV corrections to the one-loop scalar field self-energy. Continuous an wavy lines represent the scalar and photon propagators, respectively. Dashed lines respresent a LV vertex insertion.}
	\label{fig00}
\end{figure}

Let us start with the computation of the LV correction to the one-loop scalar field propagation is depicted in Figure \ref{fig00}. The total amplitude is
\begin{eqnarray}\label{phiselv01}
\Sigma(p) &=& -e^2Q_1 \int{\frac{d^Dk}{(2\pi)^D}} \Big\{ (k\cdot u)\left[ \frac{(k+p)^2}{(k^2)^2(k-p)^2}-(\xi-1)\frac{(k^2-p^2)^2}{(k^2)^2 [(k-p)^2]^2}    \right]\nonumber\\
&& -\lambda Q_1\frac{(k\cdot u)}{(k^2)^2} -e^2Q_1 
\left[ \frac{ (k-p)\cdot u}{k^2(k+p)^2}-(\xi-1)\frac{(k^2-p^2)(k+p)\cdot u}{k^2 [(k+p)^2]^2}    \right]\nonumber\\
&& +e^2Q_1 
\left[ \frac{ (k+p)\cdot u}{k^2(k-p)^2}+(\xi-1)\frac{(k^2-p^2)(k-p)\cdot u}{k^2 [(k-p)^2]^2}    \right] \Big\}.
\end{eqnarray}

Writing the integrals in terms of the integrals defined in the Appendix, we find
\begin{eqnarray}\label{phiselv02}
	\Sigma(p) &=& \frac{i\pi^2e^2 Q_1(p\cdot u)}{2(2\pi)^D}
	\Big[ 2	{\bf B}_{\{1,0\},\{1,0\}}^{(D)}  - 2 (\xi-2)	{\bf A}_{\{2,0\}}^{(D)}
\nonumber\\
&&	+ 2 p^2 (2\xi-3) {\bf B}_{\{1,0\},\{2,0\}}^{(D)}-(\xi-1)p^4 {\bf B}_{\{2,0\},\{2,0\}}^{(D)} \Big].
\end{eqnarray}

The UV divergent component of the expression can be represented as
\begin{eqnarray}\label{sec2eq01}
	\Sigma(p)=-iQ_1(p\cdot u)\left[ \frac{e^2 \left(\xi-3\right)}{16 \pi ^2 \epsilon}
	+ \delta_{Q_1}	\right].
\end{eqnarray}
\noindent Here, $\delta_{Q_1}$ stands for a counterterm, and $\epsilon=(4-D)/2$ with $D$ denoting the dimension of spacetime. Imposing finiteness to the function 	$\Sigma(p)$, we easily find 
\begin{eqnarray}\label{deltaQ1}
	\delta_{Q_1}=-\frac{e^2 \left(\xi-3\right)}{16 \pi ^2 \epsilon}.
\end{eqnarray}

\begin{figure}[ht]
	\includegraphics[angle=0,width=16cm]{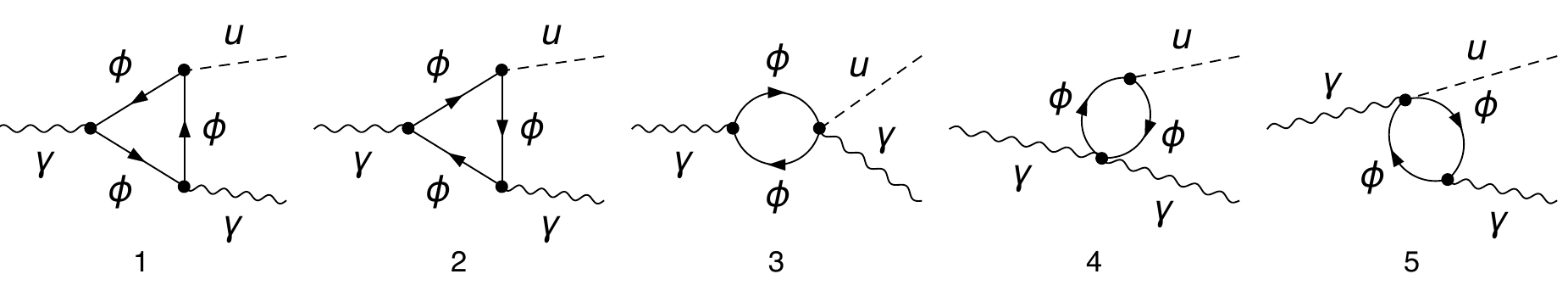}
	\caption{Feynman diagrams for the LV corrections to the one-loop photon self-energy.}
	\label{fig01}
\end{figure}

Now, let us compute the LV correction to the one-loop photon self-energy. The corresponding diagrams are illustrated in the Figure \ref{fig01}. The amplitude is given by
\begin{eqnarray}
\Pi^{\mu\alpha}(p)=e^2Q_1\int\frac{d^Dk}{(2\pi)^D}\left[ \frac{b^{\alpha} (2k-p)^\mu}{k^2(k-p)^2} +\frac{b^{\mu} (2k-p)^\alpha}{k^2(k-p)^2}+\frac{2k\cdot b g^{\mu\alpha}}{(k^2)^2} \right]=0.
\end{eqnarray}	

It is important to note that the one-loop correction to the CFJ term vanishes because in scalar QED, unlike the spinor one, there is no possibility to generate the Levi-Civita tensor. However, this scenario changes at the two-loop order, as at this level, there are insertions of the CFJ vertex in the virtual photon within the loops.

\begin{figure}[ht]
	\includegraphics[angle=0,width=16cm]{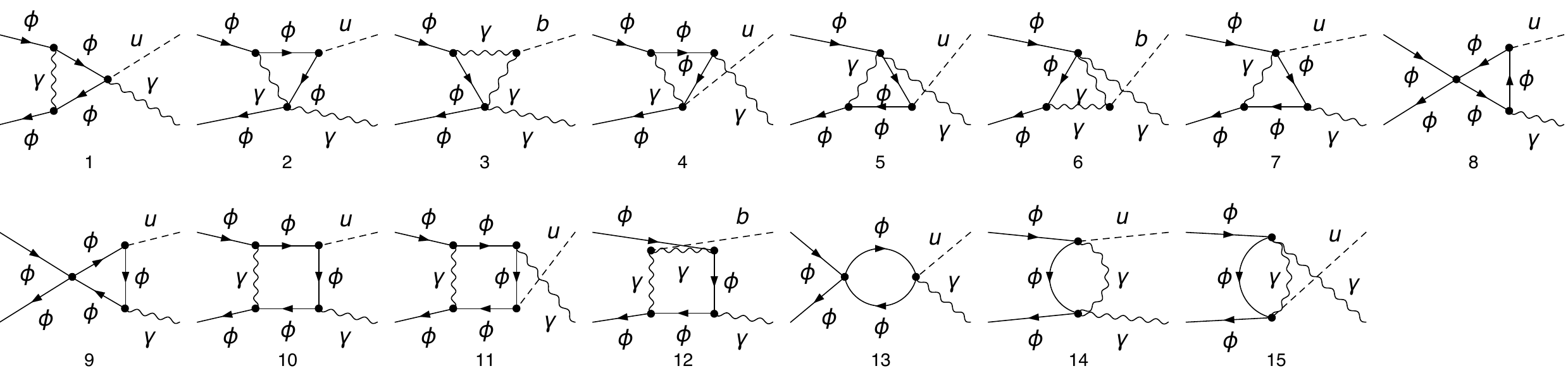}
	\caption{Feynman diagrams for the LV corrections to the one-loop photon-matter vertex interaction.}
	\label{fig2b}
\end{figure}

At last, we will calculate the one-loop LV corrections to the vertex associated with the matter-photon interaction. These diagrams are illustrated in Figure \ref{fig2b}. The corresponding UV divergent expression is as follows:
\begin{eqnarray}\label{vertexLV}
\Gamma^\mu(p) &=& -ieQ_1  u^\mu \left(\delta_{Q_1}+\frac{e^2 (\xi-3)}{16\pi^2\epsilon} +\mathrm{finite} \right).
\end{eqnarray}  

From Eqs. \eqref{phiselv01} and \eqref{vertexLV}, it becomes evident that the UV divergent component of the LV effective Lagrangian can be cast as
\begin{eqnarray}\label{eff_lag}
	\mathcal{L}_{LV}&=&  \frac{i}{2}\left(1-\frac{e^2(\xi-3)}{16\pi^2\epsilon}-\delta_{Q_1} \right) Q_1 u^{\mu}\left[ \phi^* D_\mu\phi-\phi  (D_{\mu}\phi)^*\right].
\end{eqnarray}
We see that this result perfectly reproduces the form of the LV term in our Lagrangian, confirming therefore the multiplicative renormalizability of our theory. The similar situation occurs in the CPT-even LV scalar QED as well \cite{our2loop}.


\subsection{Calculation of the  LV corrections to the  two-loop photon self-energy} \label{sec3-twoloop}

\begin{figure}[ht]
	\includegraphics[angle=0 ,width=14cm]{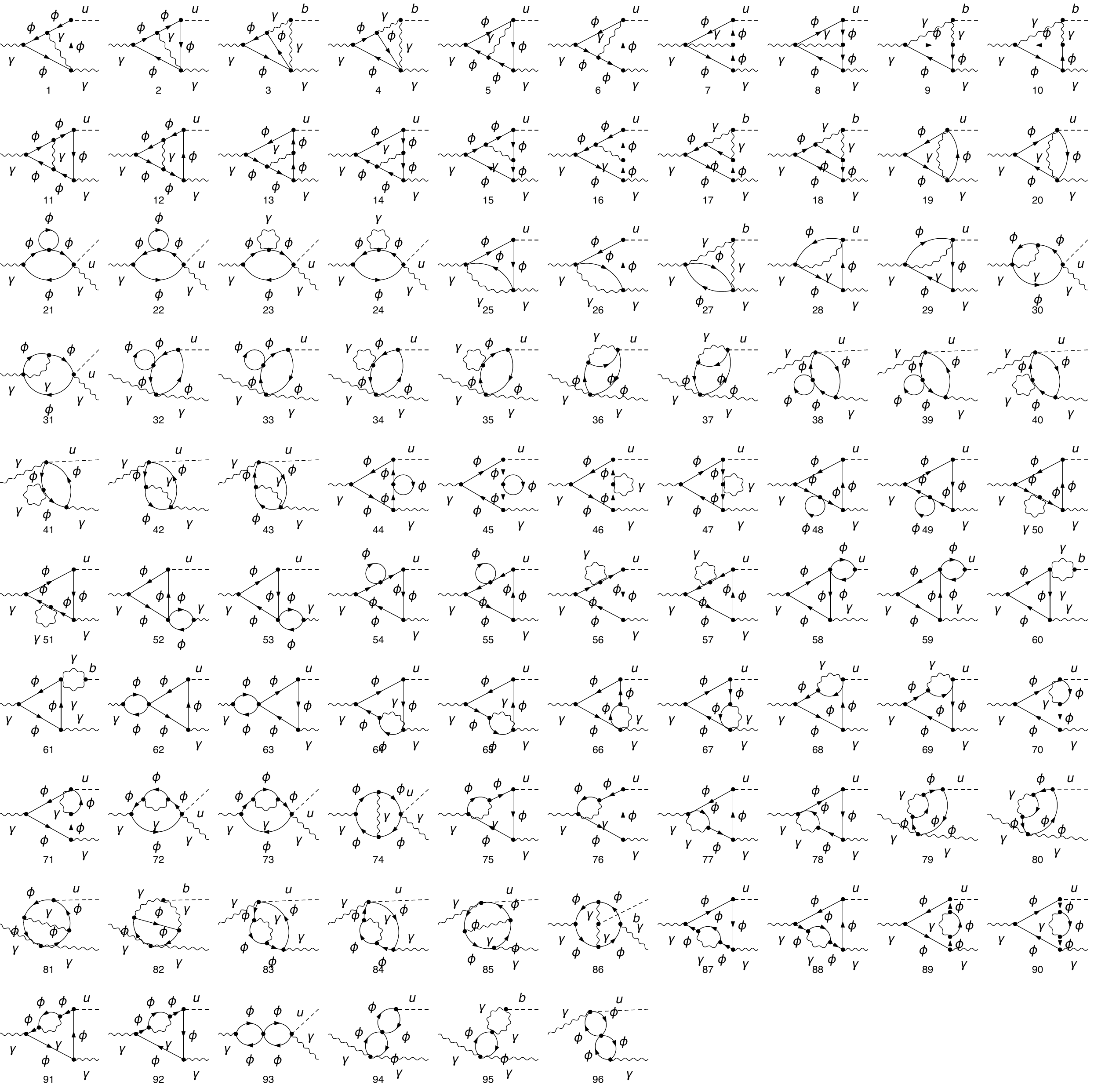}
	\caption{Feynman diagrams for the LV corrections to the two-loop photon self-energy.}
	\label{fig02}
\end{figure}

\begin{figure}[ht]
	\includegraphics[angle=0 ,width=14cm]{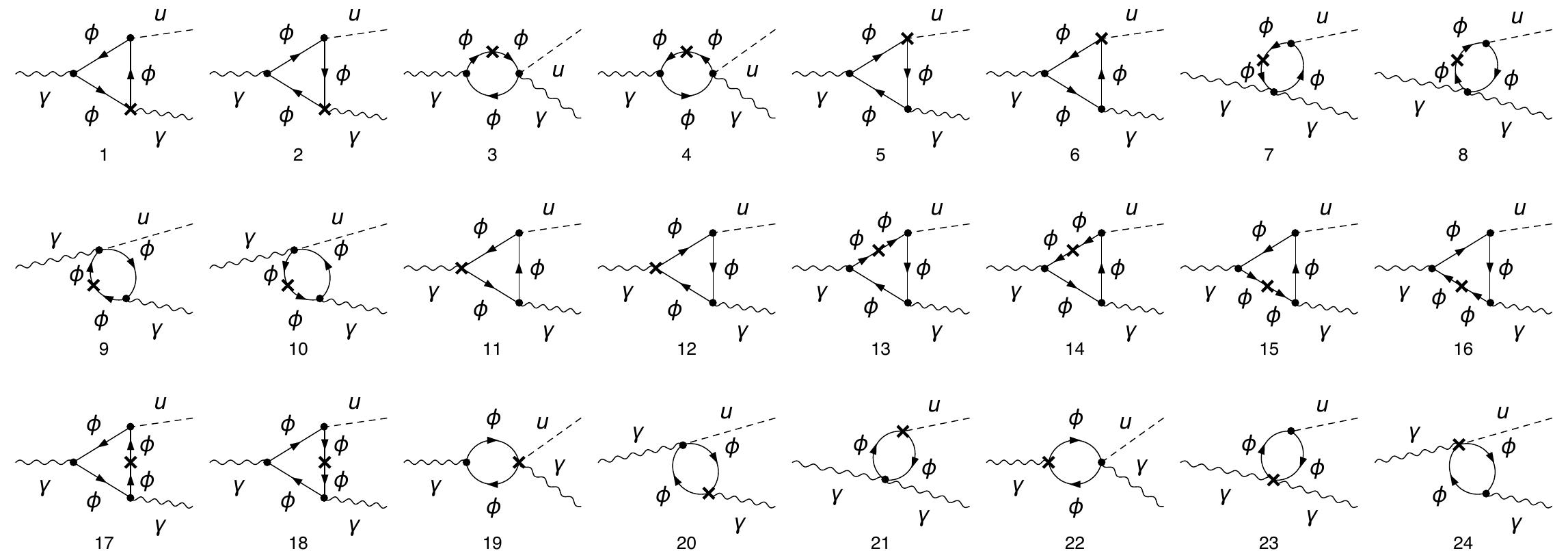}
	\caption{These one-loop Feynman diagrams illustrate the LV corrections to the photon self-energy, featuring a counterterm insertion.}
	\label{fig03}
\end{figure}

The LV corrections to the two-loop photon self-energy are depicted in Figure \ref{fig02}. While the diagrams shown in Figure \ref{fig03} are expected to contribute to the photon self-energy at the same order as the two-loop diagrams, they collectively cancel each other out, resulting in a net contribution equal to zero. Upon combining all the contributions from Figure \ref{fig02}, we obtain the following result:
\begin{eqnarray}
	\Pi^{\mu\beta}_{2l}(p)&=&  \frac{8e^4Q_2 }{(D-4)^2(D-1)p^2}b_{\sigma}  p_{\rho}  \epsilon^{\beta \mu  \rho \sigma}\Big[ (D-4)p^2 (\textbf{B}^{(D)}_{\{1,0 \},\{1,0 \}} )^2 \nonumber\\
	&& +(D^4-12 D^3+49 D^2-82 D+48) \textbf{J}^{(D)}_{\{1,0 \},\{1,0 \},\{1,0 \}} \Big]
	\nonumber\\
	&=&
	\frac{e^4 Q_2 (6 \zeta (3)-5) p_{\rho} b_{\sigma} {\epsilon }^{\beta \mu  \rho \sigma}}{192 \pi^4}
\end{eqnarray}
\noindent where the definitions of the integrals are provided in the Appendix. It is worth noting that this contribution is UV finite and ambiguity-free. This constitutes the two-loop corrections to the CFJ term.

\subsection{LV corrections to the  two-loop scalar field self-energy} \label{sec4-twoloop-scalar}

\begin{figure}[ht]
	\includegraphics[angle=0 ,width=14cm]{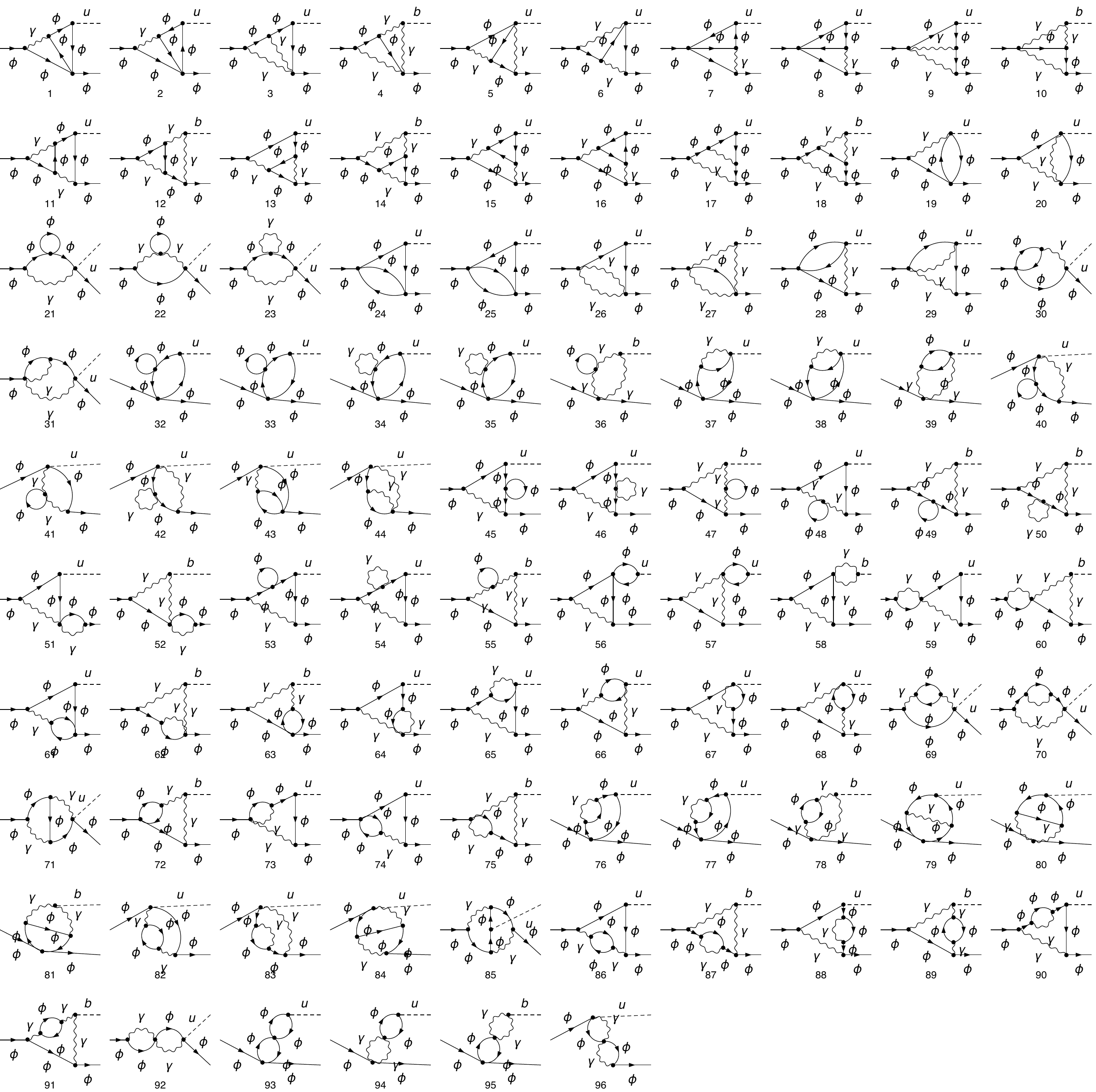}
	\caption{Feynman diagrams for the LV corrections to the two-loop scalar field self-energy.}
	\label{fig02b}
\end{figure}

\begin{figure}[ht]
	\includegraphics[angle=0 ,width=14cm]{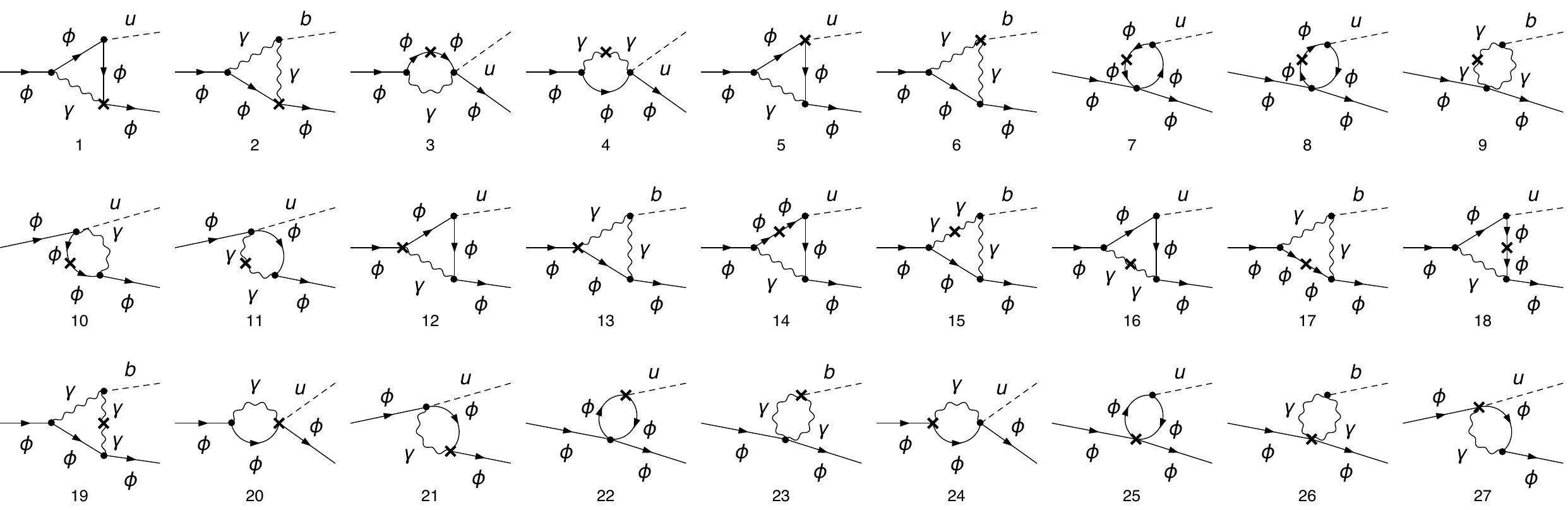}
	\caption{These one-loop Feynman diagrams illustrate the LV corrections to the scalar field self-energy, featuring a counterterm insertion.}
	\label{fig03b}
\end{figure}

The LV corrections to the two-loop scalar field self-energy are depicted in Figure \ref{fig02b}. While the diagrams shown in Figure \ref{fig03b} are expected to contribute to the same process at the same order as the two-loop diagrams.Adding all the contributions from Figure \ref{fig02b}, we obtain the following result:
\begin{eqnarray}
	\Sigma_{2l}(p)&=&- \frac{iQ_1(p\cdot u)}{2048\pi^4} \Big{\{}
	\frac{4(\xi^2-6\xi+10)e^4}{\epsilon^2}+\frac{1}{\epsilon}\Big[\lambda^2
	-8e^4((\xi-6)\xi+10)\ln{\left(-\frac{p^2}{4\pi\mu^2}\right)}\nonumber\\
	&&-8e^4 ((\xi-2)\xi+\gamma_E((\xi-6)\xi+10)-5) \Big]
	\Big{\}},
\end{eqnarray}
\noindent where $\gamma_E$ is the Euler-Mascheroni constant and $\mu$ is a mass scale introduced by the regularization.

Furthermore, when we combine all the contributions from Figure \ref{fig03b}, we find
\begin{eqnarray}
	\Sigma_{2l-CT}(p)&=&\frac{iQ_1e^2(p\cdot u)}{16\pi^2} \Big{\{}
\delta_3\Big[ \frac{3}{\epsilon}+3\ln\left(-\frac{p^2}{4\pi\mu^2}\right) +3\gamma_E-1\Big]\nonumber\\
&& +\delta_{Q_1}\Big[- \frac{\xi-3}{\epsilon}+\gamma_E(\xi-3)+\xi+1+(\xi-3) \ln\left(-\frac{p^2}{4\pi\mu^2}\right)
\Big]
	\Big{\}},
\end{eqnarray}
\noindent where $\delta_{Q_1}$ is the counterterm as defined in Eq. \eqref{deltaQ1}, and $\delta_3=-\frac{e^2}{48\pi^2\epsilon}$ represents the conventional photon wave-function renormalization \cite{brito2}.

Upon combining all the contributions, including counterterms, one- and two-loop contributions, and substituting the expressions for $\delta_3$ and $\delta_{Q_1}$, we obtain
\begin{eqnarray}
	\Sigma_{LV}(p)&=& -iQ_1 (p\cdot u)\Big{\{}  \delta_{Q_1} +
	 \frac{ e^2\left(\xi-3\right)}{16 \pi ^2 \epsilon} 
	  -\frac{e^4	(\xi^2-6\xi+10)}{512\pi^4 \epsilon^2} 
	+\frac{(3\lambda^2+40 e^4)}{6144\pi^4\epsilon}+\mathrm{finite}
	\Big{\}}.	
\end{eqnarray}
\noindent The counterterm $\delta_{Q_1}$ is determined up to two-loop order by imposing the requirement of UV finiteness on the scalar field self-energy.

\section{Final Remarks}\label{summary}
 

In this paper, we have undertaken a comprehensive exploration of one- and two-loop contributions within the context of CPT-odd LV QED. Our calculations encompass one-loop corrections to the LV vertices of the model, the two-loop LV corrections affecting the UV divergent aspect of the scalar field self-energy, and the two-loop contribution to the CFJ term. Of particular significance is our explicit demonstration of the two-loop finiteness of the CFJ term. It is worth noting that, in the one-loop approximation, this term remains finite in the presence of a nontrivial vacuum, becoming zero if the vacuum expectation value (v.e.v.) of a scalar field is zero \cite{Scarp2013}.

This finding implies the potential for the finiteness of the CFJ term in any CPT-odd LV QED, regardless of the nature of the matter coupled to the gauge field. It is essential to highlight that, in contrast to spinor QED, our theory ensures the absence of ambiguities associated with the CFJ term, at least when applying the Passarino-Veltman prescription. This arises naturally from the fact that our theory does not involve Dirac matrices, whose definition in a $D$-dimensional space-time can be ambiguous, particularly for non-integer values of $D$. Consequently, our work has offered a definitive and ambiguity-free method to generate a correction to the CFJ term without relying on a nontrivial vacuum.

The possible continuation of this study could consist in an investigation of the Higgs mechanism in the two-loop approximation, i.e. in introducing the spontaneous breaking of $U(1)$ gauge symmetry, and, probably, in higher-loop studies.  We expect to pursue these aims in forthcoming papers.  


\acknowledgments

The work of A. Yu.\ P. has been partially supported by the CNPq project No. 301562/2019-9. The work of A. C. L. has been partially supported by the CNPq project No. 404310/2023-0.

\appendix*

\section{Definition of the integrals}

The basic integrals used in this text and employed by the TARCER package, as defined in Ref. \cite{tarcer}, are
\begin{eqnarray}
	{\bf A}_{\{n,0\}}^{(D)} \!\! &=& \!\! \frac{1}{\pi^{D/2}}\int\frac{d^Dk}{(k^2)^n};\\
	{\bf B}_{\{n,0\},\{m,0\}}^{(D)} \!\! &=& \!\! \frac{1}{\pi^{D/2}}\int \frac{d^Dk}{(k^2)^n[(k-p)^2]^m};\\
	{\bf J}_{\{1,0\},\{1,0\},\{1,0\}}^{(D)} \!\! &=& \!\! \frac{1}{\pi^{D}}\int\frac{d^Dk_1~d^Dk_2}{k_1^2(k_1-k_2)^2(k_2-p)^2},
\end{eqnarray}
\noindent where $D=4-2\epsilon$. The one-loop integrals are well-known and have been extensively studied in the literature. A comprehensive calculation of the two-loop integral was carried out in Ref. \cite{tsil}.

\newpage

\end{document}